\documentclass[sigconf]{acmart}

\AtBeginDocument{%
  \providecommand\BibTeX{{%
    \normalfont B\kern-0.5em{\scshape i\kern-0.25em b}\kern-0.8em\TeX}}}

\setcopyright{acmcopyright}
\copyrightyear{2021}
\acmYear{2021}

\acmConference[ICOOOLPS '21]{ICOOOLPS '21}{July 13, 2021}{Online}
\acmBooktitle{ICOOOLPS '21, June 13, 2021, Online}

\usepackage{acro}       
\usepackage{calc}       
\usepackage{csquotes}   
\usepackage{listings}   
\usepackage{tikz}       

\usetikzlibrary{calc, positioning, decorations.pathreplacing}
\makeatletter
\newlength{\linenumwidth} 
\newlength{\numwidth}%
\def\lst@PlaceNumber{%
  \makebox[\numwidth+1em][l]{%
    \makebox[\numwidth][r]{\normalfont\lst@numberstyle{\thelstnumber}}%
  }%
}
\makeatother

\lstdefinelanguage{mvsl} {
  keywords    = [1]{struct, func, let, var, in, inout},
  keywords    = [2]{Int, Float},
  morecomment = [l]{//},
  morecomment = [s]{/*}{*/},
  morestring  = [b]",
}

\lstdefinelanguage{swift} {
  keywords    = [1]{class, struct, protocol, extension, associatedtype, override, static, init, func, return, let, var, inout, where, in, as, async, await},
  keywords    = [2]{super, self, Self},
  morecomment = [l]{//},
  morecomment = [s]{/*}{*/},
  morestring  = [b]",
}

\DeclareAcronym{mvs}{
  short = MVS,
  long  = mutable value semantics,
}
\DeclareAcronym{pds}{
  short = PDS,
  long  = passive data structure,
}

\newcommand{\mvslang}{MVSL}

\begin{document}

\title{Native Implementation of Mutable Value Semantics}

\author{Dimitri Racordon}
\email{dimitri.racordon@unige.ch}
\orcid{0000-0003-0299-3993}
\affiliation{
  \department{Faculty of Science}
  \institution{University of Geneva}
  \country{Switzerland}
}

\author{Denys Shabalin}
\email{shabalin@google.com}
\affiliation{
  \institution{Google}
  \country{Switzerland}
}

\author{Daniel Zheng}
\email{danielzheng@google.com}
\affiliation{
  \institution{Google}
  \country{United States}
}

\author{Dave Abrahams}
\email{dabrahams@google.com}
\affiliation{
  \institution{Google}
  \country{United States}
}

\author{Brennan Saeta}
\email{saeta@google.com}
\affiliation{
  \institution{Google}
  \country{United States}
}

\renewcommand{\shortauthors}{Trovato and Tobin, et al.}

\begin{abstract}
Unrestricted mutation of shared state is a source of many well-known problems.
The predominant safe solutions are pure functional programming, which bans mutation outright, and flow sensitive type systems, which depend on sophisticated typing rules.
Mutable value semantics is a third approach that bans sharing instead of mutation, thereby supporting part-wise in-place mutation and local reasoning, while maintaining a simple type system.
In the purest form of mutable value semantics, references are second-class: they are only created implicitly, at function boundaries, and cannot be stored in variables or object fields.
Hence, variables can never share mutable state.

Because references are often regarded as an indispensable tool to write efficient programs, it is legitimate to wonder whether such a discipline can compete other approaches.
As a basis for answering that question, we demonstrate how a language featuring mutable value semantics can be compiled to efficient native code.
This approach relies on stack allocation for static garbage collection and leverages runtime knowledge to sidestep unnecessary copies.
\end{abstract}




\maketitle

\section{Introduction}

Software development continuously grows in complexity, as applications get larger and hardware more sophisticated.
One well-established principle to tackle this challenge is \emph{local reasoning}, descrbibed by \citet{DBLP:conf/csl/OHearnRY01} as follows:
\blockquote{\itshape
To understand how a program works, it should be possible for reasoning and specification to be confined to the cells that the program actually accesses.
The value of any other cell will automatically remain unchanged.}

There are two common ways to uphold local reasoning.
One takes inspiration from pure functional languages and immutability.
Unfortunately, this paradigm may fail to capture the programmer's mental model, or prove ill-suited to express and optimize some algorithms~\cite{DBLP:journals/jfp/ONeill09}, due to the inability to express in-place mutation.

Another approach aims to tame aliasing.
Newer programming languages have successfully blended ideas from ownership types~\cite{DBLP:series/lncs/ClarkeOSW13}, type capabilities~\cite{DBLP:conf/ecoop/HallerO10}, and region-based memory management~\cite{DBLP:journals/lisp/TofteBEH04} into flow-sensitive type systems, offering greater expressiveness and giving more freedom to write efficient implementations.
Unfortunately, these approaches have complexity costs that significantly raise the entry barrier for inexperienced developers~\cite{turner:rust-survey}.

\Ac{mvs} offers a tradeoff that does not add the complexity inherent to flow-sensitive type systems, yet preserves the ability to express in-place, part-wise mutation.
It does so treating references as a ``second-class'' concept.
References are only created at function boundaries by the language implementation, and only if the compiler can prove their uniqueness.
Further, they can neither be assigned to a variable nor stored in object fields.
Hence, all values form disjoint topological trees, whose roots are assigned to the program's variables.

The reader may understandably worry about expressiveness and efficiency, as references are often held as indispensable for both aspects.
We note that a large body of software projects already address expressiveness concerns empirically, such as the Boost Graph Library\cite{bgl-book}, leveraging \ac{mvs} to elucidate recurring questions surrounding equality, copies, and mutability, and develop generic data structures and algorithms~\cite{fm2gp}.

This paper focuses on the question of efficiency.
We discuss an approach for compiling languages featuring \ac{mvs} to native code, relying on stack allocation for static garbage collection and using runtime information to elide unnecessary copies.
We present it in the context of a toy programming language, called \mvslang{}, inspired by Swift, for which we have written a compiler.
Our implementation is available as an open-source project hosted on GitHub: \url{https://github.com/kyouko-taiga/mvs-calculus}.

\section{A quick tour of MVSL}
\label{sec:mvsl}

\mvslang{} is a statically typed expression-oriented language, designed to illustrate the core principles of \ac{mvs}.
In \mvslang{}, a program is a sequence of structure declarations, followed by a single expression denoting the entry point (i.e., the \lstinline|main| function in a C program).

A variable is declared with the keyword \lstinline|var| followed by a name, a type annotation, an initial value, and the expression in which it is bound.
A constant is declared similarly, with the keyword \lstinline|let|.
\begin{lstlisting}
var foo: Int = 4 in
let bar: Int = foo in bar
\end{lstlisting}

There are three built-in data types in the MVSL:
\lstinline|Int| for signed integer values, \lstinline|Float| for floating-point values, and a generic type \lstinline|[T]| for arrays of type \lstinline|T|.
In addition, the language supports two kinds of user-defined types: functions and structures.
A structure is a heterogeneous data aggregate, composed of zero or more fields.
Each field is typed explicitly and associated with a mutability qualifier (\lstinline|let| or \lstinline|var|) that denotes whether it is constant or mutable.
\begin{lstlisting}
struct Pair {
  var fs: Int; var sn: Int
} in
var p: Pair = Pair(4, 2) in p
\end{lstlisting}
Fields of a structure can be of any type, but type definitions cannot be mutually recursive.
Hence, all values have a finite representation.

All types have value semantics.
Thus, all values form disjoint topological trees rooted at variables or constants.
Further, all assignments copy the right operand and never create aliases, departing from the way aggregate data types typically behave in popular object-oriented programming languages, such as Python or Java.
\begin{lstlisting}
struct Pair { ... } in
var p: Pair = Pair(4, 2) in
var q: Pair = p in
q.sn = 8 in
p // p is equal to Pair(4, 2)
  // q is equal to Pair(4, 8)
\end{lstlisting}

Immutability applies transitively.
All fields of a data aggregate assigned to a constant are also treated as immutable by the type system, regardless of their declaration.
\begin{lstlisting}
struct Pair { ... } in
let p: Pair = Pair(4, 2) in
p.sn = 8 in p // <- type error
\end{lstlisting}
Likewise, all elements of an array are constant if the array itself is assigned to a constant.
\begin{lstlisting}
struct Pair { ... } in
let a: [Pair] = [Pair(4,2), Pair(5,3)] in
a[0].sn = 8 in a // <- type error
\end{lstlisting}

Functions are declared with the keyword \lstinline|func| followed by a list of typed parameters, a codomain, and a body.
Functions are anonymous but are first-class citizen values that can be assigned, passed as an argument, or returned from other functions.
Arguments are evaluated eagerly and passed by copy.
Further, functions are allowed to \emph{capture} identifiers from their declaration environment.
Such captures also result in copies and stored in the function's closure, thus preserving value independence.
\begin{lstlisting}
var foo: Int = 42 in
var fn: () -> Int {
  foo = foo + 1 in foo
} in
let bar = fn() in
bar // foo is equal to 0
    // bar is equal to 1
\end{lstlisting}

To implement part-wise in-place mutation across function boundaries, values of parameters annotated \lstinline|inout| can be mutated by the callee.
At an abstract level, an \lstinline|inout| argument is copied when the function is called and copied back to its original location when the function returns.\footnote{The Fortran enthusiast may think of the so-called ``call-by-value/return'' policy.}
At a more operational level, an \lstinline|inout| argument is simply passed by reference.
Of course, \lstinline|inout| extends to multiple arguments, with one important restriction: overlapping mutations are prohibited to prevent any writeback from being discarded.
\begin{lstlisting}
struct Pair { ... } in
struct U {} in
let swap: (inout Int, inout Int) -> U
  = (a: inout Int, b: inout Int) -> U {
    let tmp = a in
    a = b in
    b = tmp in U()
  } in
var p = Pair(4, 2) in
_ = swap(&p.fs, &p.sn)
in p // p is equal to Pair(2, 4)
\end{lstlisting}

A more exhaustive specification of \mvslang{}, as well as more elaborate program examples, are available in the GitHub repository.

\section{Native implementation}

This section describes the strategy we implemented to compile \mvslang{} to native code.

\subsection{Memory representation}

\lstinline|Int| and \lstinline|Float| are built-in numeric types that typically have a 1-to-1 correspondence with machine types.
Since \lstinline|struct| definitions cannot be mutually recursive, all values of a structure have a finite memory representation (more on that later).
Therefore, they can be represented as \ac{pds}, where each field is laid out in a contiguous fashion.

In the absence of first-class references, it is fairly easy to identify the lifetime of a value: it begins when the value is assigned to a variable and ends when said variable is reassigned or goes out of scope.
Following this observation, an obvious choice to handle memory is to rely on stack allocation, to automate memory management.

A type is \emph{trivial} if it denotes a number or a composition of trivial types (e.g., a pair of \lstinline|Int|s).
A variable of a trivial type represents a single memory block allocated on the stack, which does not involve any particular operation to be initialized or deallocated.

Non-trivial types require more attention.
In \mvslang{}, arrays and closures require dynamic allocation, because the compiler is in general incapable of determining the number of elements in an array or the size of a closure from their signatures.
They can be represented as fixed-size data aggregates that point to heap-allocated memory, nonetheless.
Furthermore, the aforementioned observation about lifetimes remains.
Hence, the compiler can generate code to reclaim dynamically allocated memory when variables holding arrays or functions go out of scope or are reassigned.



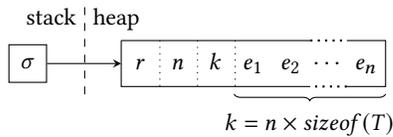
\begin{figure}
  \centering
  \begin{tikzpicture}[>=stealth]
    \tikzstyle{cell}=[minimum height=0.5cm, minimum width=0.5cm]

    \node[draw, cell] (a) {$\sigma$};

    \node[cell] (r) at ($(a)+(1.5,0)$) {$r$\strut};
    \node[cell] (n) at ($(r)+(0.5,0)$) {$n$\strut};
    \node[cell] (k) at ($(n)+(0.5,0)$) {$k$\strut};

    \node[cell] (e1) at ($(k) +(0.5,0)$) {$e_1$\strut};
    \node[cell] (e2) at ($(e1)+(0.5,0)$) {$e_2$\strut};
    \node[cell] (ei) at ($(e2)+(0.5,0)$) {$\cdots$\strut};
    \node[cell] (en) at ($(ei)+(0.5,0)$) {$e_n$\strut};

    \draw[->] (a) -- (r);

    \draw (r.north west) -- (e2.north east);
    \draw[dotted, thick] (e2.north east) -- (en.north west);
    \draw (en.north west)
      -- (en.north east)
      -- (en.south east)
      -- (en.south west);
    \draw[dotted, thick] (en.south west) -- (e2.south east);
    \draw (e2.south east) -- (r.south west) -- (r.north west);

    \draw[dotted] (r.north east) -- (r.south east);
    \draw[dotted] (n.north east) -- (n.south east);
    \draw[dotted] (k.north east) -- (k.south east);

    \draw [decorate, decoration={brace, mirror, raise=0.1cm}] (e1.south west) -- (en.south east) node [midway, yshift=-0.5cm] {$k = n \times \mathit{sizeof}(T)$};

    \draw[dashed] ($(a.east)+(0.5,0.75)$) -- ($(a.east)+(0.5,-0.5)$);
    \node[anchor=east, minimum height=0.5cm] at ($(a.east)+(0.5,0.6)$) {stack\strut};
    \node[anchor=west, minimum height=0.5cm] at ($(a.east)+(0.5,0.6)$) {heap\strut};
  \end{tikzpicture}
  \caption{In-memory representation of an array of $T$}
  \label{fig:array-repr}
\end{figure}

An array is represented by a pointer $\sigma$ to a contiguous block of heap-allocated memory.
The block is structured as a tuple $\langle r, n, k, \overline{e} \rangle$ where $r$ is a reference counter, $n$ denotes the number of elements in the array, $k$ denotes the capacity of the array's payload (i.e., the size of its actual contents) and $\overline{e}$ is a payload of $k$ bytes.
The counter $r$ serves to implement the so-called \emph{copy-on-write} optimization (see Section~\ref{sec:cow}).
Figure~\ref{fig:array-repr} depicts the in-memory representation of an array of elements of some type $T$.

The capacity $k$ of an array is typically different than of number of its elements $n$, because the former depends on the size an element in memory.
For example, an array of 16-bit integer values $[42, 1337]$ can be represented by a tuple $\langle 1, 2, 4, 42, 0, 5, 57 \rangle$ (assuming a little-endian system).
The array contains two elements, thus $n=2$, yet its capacity $k=4$, since each element occupies two bytes.

Closures use a \ac{pds} $\langle \phi, \epsilon, c, d \rangle$ where $\phi$ is a pointer to a function implementing the closure, $\epsilon$ is a pointer to the closure's environment, and $c$ and $d$ are pointers to synthesized routines that respectively copy and destroy the closure (see Section~\ref{sec:metatypes}).

The function pointed by $\phi$ is obtained by \emph{defunctionalizing}~\cite{reynolds:1998:higher-order-programming} the closure.
This process transforms the closure into a global function in which all captured identifiers are lifted into an additional parameter for the closure's environment.

\subsection{Copying and destroying values}
\label{sec:metatypes}

Recall that assignments result in copies of their right operand.
Since \mvslang{} is a statically typed language, the compiler knows \emph{how} to copy values of fixed size.
For trivial types, the operation consists of a mere bitwise copy of the right operand.

The situation is a bit more delicate for non-trivial types.
For arrays, a first issue is that the size of its heap-allocated storage cannot be determined statically.
Instead, it depends on the value of $k$ in the \ac{pds} that represents the array.
A second issue is that copying may involve additional operations if the elements contained in the array are dynamically sized as well.
In this case, a bitwise copy would improperly create aliases on the heap-allocated memory, breaking value independence.
Instead, each element must be copied individually, allocating new memory as necessary.

One solution is to synthesize a function for each data type that is applied whenever a copy should occur.
If the type is trivial (i.e., it does not involve any dynamic allocation), then this function is equivalent to a bitwise copy.
Otherwise, it implements the appropriate logic, calling the copy function of each contained element.

Similarly, the logic implementing the destruction of a value can be synthesized into a destructor.
If the type is trivial, then this destructor is a no-op.
Otherwise, it recursively calls the destructor of each contained element and frees the memory allocated for all values being destroyed.

\subsection{Crossing function boundaries}

At function boundaries, \acp{pds} are exploded into scalar arguments and passed directly through registers, provided the machine has enough of them.
If the structure is too large, it is passed as a pointer to a stack cell in the caller's context, in which a copy of the argument while have been stored before the call.

An \lstinline|inout| argument is passed as a (possibly interior) pointer.
If it refers to a local variable or one of its fields, then it is passed as a pointer to the stack.
If it refers to the element of an array, then it is passed as a pointer within the array's storage.

Note: the compiler can guarantee that the pointee can never be outlived, because the language disallows the pointer to escape in any way.
In fact, the value of the pointer itself is not accessible.
The callee can only dereference it, either to store or load a value.
Further, recall that the type system guarantees exclusive access to any memory location.
Hence, pointers representing \lstinline|inout| arguments are known to be unique.

\subsection{Avoiding unnecessary copies}
\label{sec:optimizations}

The implementation we have described so far generates a fair amount of memory traffic, as copies are created every time a value is assigned to a variable or passed as an argument.
Much of this traffic is unnecessary, though, because most original values are destroyed immediately after being copied, or because copied values might never be mutated and could have been shared.
We now briefly discuss three techniques to eliminate unnecessary copies.

\subsubsection{Move semantics}

A recurring pattern is to assign values just after they have been created.
For example, consider the expression \lstinline|let x: [Int] = [1, 2] in f(x)|.
The value of the array is assigned directly after its creation.

A naive implementation will evaluate the right operand, resulting in the creation of a new array value, copy this value to assign \lstinline|x| and then destroy the original.
Clearly, the copy is useless, since the original value will never be used.
Hence, one can \emph{move} the temporary value into the variable rather than copying it.

Moving a value boils down to a bitwise copy.
We said earlier that such a strategy was incorrect in the case of an array because it would create aliases.
In this particular case, however, the other alias is discarded immediately and therefore the variable remains independent once the assignment is completed.

A similar situation occurs when arguments are being copied.
In the above expression, \lstinline|x| must be copied before it is passed as an argument to the function \lstinline|f|.
However, because the remainder of the expression does not mention \lstinline|x| anymore, this copy can be elided and the value of \lstinline|x| can be \emph{moved} into the function.

\subsubsection{Copy-on-write}
\label{sec:cow}

Copies of immutable values to immutable bindings can obviously be elided.
Indeed, aliasing is harmless in the absence of mutation, and we can \emph{simulate} value semantics on top of shared immutable states.
In contrast, assigning a mutable value to an immutable binding or vice versa typically requires a copy, because the value might be mutated later.
Similarly, assigning a mutable value to a mutable binding also requires a copy.

Nonetheless, it is possible that neither the original nor the copy end up being actually mutated, perhaps because the mutation depends on a condition that is evaluated at runtime.
In this case, unfortunately, the compiler must conservatively assume that a mutation will occur and perform a copy to preserve value independence.

One simple mechanism can be used to workaround this apparent shortcoming: \emph{copy-on-write}.
Copy-on-write leverages runtime knowledge to delay copies until they are actually needed.
Heap-allocated storage is associated with a counter that keeps track of the number of pointers to that storage.
Every time a value is copied, an alias is created and the counter is incremented.
The value of this counter is checked when mutation actually occurs, at runtime.
If it is greater than one, the counter is decremented, the storage is duplicated and the mutation is performed on a copy.
Otherwise, the mutation is performed on the original.

The counter is decreased whenever the destructor of a value referring to the associated storage is called.
If it reaches zero, then the contents of the storage are destroyed and deallocated.

\subsubsection{Leveraging local reasoning}

We cited \citet{DBLP:conf/csl/OHearnRY01} in the introduction to emphasize the importance of local reasoning for human developers.
We add that local reasoning is also an invaluable tool for automated program optimizations, as it eliminates the need for conservative assumptions about the use of memory.
In particular, one can easily identify and discard irrelevant mutations, because one can assume those cannot be observed elsewhere.

\begin{lstlisting}
struct Pair { ... } in
var p: Pair = Pair(4, 2) in
let q: Pair = p in
p.fs = 8 in
Pair(p.sn, q.fs).
\end{lstlisting}

Consider the above program.
Thanks to local reasoning, an optimizer can safely discard the assignment to \lstinline|p.fs| at line 4, because its effect is never observed.
Without this assignment, it becomes clear that \lstinline|p| and \lstinline|q| are the exact same value, and the former's copy can be elided.
Eventually, constant propagation will deduce that the program is equivalent to the expression \lstinline|Pair(2, 4)|.

Note: such optimizations are fairly standard in off-the-shelf optimizers.
Our own implementation simply relies on the default optimization passes of LLVM~\cite{DBLP:conf/cgo/LattnerA04}.

\section{Managed environments}

We did not discuss any strategy to execute \mvslang{} in managed runtime environments, where stack allocation and interior pointers are typically unavailable.
We note that the strategies we have presented in Section~\ref{sec:optimizations} are applicable nonetheless.
Move assignments can be substituted by merely copying references, copy-on-write can operate similarly in a managed environment and local reasoning enables the same kind of optimizations.

In addition, copies of large immutable structures can be avoided by memoizing them in a uniqueness table, segregated by data types for efficient lookup.
Intuitively, memoization should be particularly beneficial for programs that often test for equality.

One important challenge relates to \lstinline|inout| arguments.
A naive solution consists of boxing every field and every element into a distinct object.
Unfortunately, this approach should be likely inefficient, due to the loss of cache locality.
A cleverer strategy could represent \lstinline|inout| parameters as writeable keypaths (i.e., closures allowing write access to a specific path in a data structure).
We leave further investigation on that front to future work.

\section{Conclusion}

We present an approach to compile programming languages featuring \acl{mvs} into native code.
We rely heavily on stack allocation to implement static garbage collection, and insert calls to synthesized destructors to deallocate dynamically-sized values automatically.
Furthermore, we leverage \emph{copy-on-write} to elide unnecessary memory traffic at runtime.



\bibliographystyle{ACM-Reference-Format}
\bibliography{main}


\end{document}